\documentstyle[fleqn,espcrc2]{article}

\def\beq{\begin{equation}}
\def\eeq{\end{equation}}
\def\bea{\begin{array}}
\def\eea{\end{array}}
\def\beqn{\begin{eqnarray}}
\def\eeqn{\end{eqnarray}}
\newcommand{\eq}[1]{(\ref{#1})}

\newcommand{\Z}{{Z \!\!\! Z}}
\newcommand{\dd}{\mbox{d}}
\newcommand{\dD}{{\cal D}}
\newcommand{\cZ}{{\cal Z}}

\newcommand{\cL}{{\cal L}}

\newcommand{\dual}[1]{{#1}^d}
\newcommand{\diff}{\partial}

\def\u1{{U(1)}}
\def\su2{{SU(2)}}

\title{Short strings and gluon propagator in the infrared region.
\vskip-3cm
\leftline{\small ITEP-TH-41/99}
\vskip2.6cm
}
\author{M.N.~Chernodub\address{Institute of Theoretical and
Experimental Physics,\\B.Cheremushkinskaya 25, Moscow, 117259, Russia},
F.V.~Gubarev${}^{\rm a}{}$, M.I.~Polikarpov${}^{\rm a}{}$,
V.I.~Zakharov\address{Max-Planck Institut f\"ur Physik,\\
F\"ohringer Ring 6, 80805 M\"unchen, Germany}\thanks{Material based on talks
by F.V.~Gubarev and M.I.~Polikarpov at Lattice '99, Pisa, Italy.}}

\begin{document}
\begin{abstract}
We discuss how infrared region influence on short distance physics via new
object, called ``short string''. This object exists in confining theories and violates
the operator product expansion. Most analytical results are obtained for the dual
Abelian Higgs theory, while phenomenological arguments are given for QCD.
\end{abstract}
\maketitle
\section{Introduction}

Below we describe effects produced by the recently introduced object, short
string. Short strings \cite{GuPoZa98} are responsible for non-standard
power corrections in QCD, which are important in the ultraviolet region. On
the other hand short strings exist due to confinement and in this sense
they are induced by large distance physics. Short strings are not
point-like objects and violate such fundamental principle as operator
product expansion (OPE).

It is interesting to find the short string effects in QCD, but
confinement in QCD can not be described from first principles and at the
beginning (in Sects. 2-4) we consider the dual Abelian Higgs model (AHM). In
this model electric charges are confined at the classical level, and it is
simple enough to study both large and small distances. Moreover, due to the
Abelian dominance \cite{AbDom}, the dual AHM can be considered as the
effective infrared theory for QCD. At small distances dual AHM is
not related to QCD, and we consider it to study general properties of short 
strings. We show that short strings give rise to special singularities in
the photon propagator and produce the linear correction, $\sigma_0 r$, to
the Coulomb potential, $\alpha/r$, at small distances.

In Sect. 5 we discuss numerical (lattice calculations) and phenomenological
facts supporting the existence of short strings in QCD.

\section{Short strings in classical equations of motion for AHM}
Consider the following problem: what are corrections to Coulomb
potential at small distances in QCD? If we neglect the trivial
perturbative dependence of the bare coupling on $r$, we have in the standard
approach (perturbation theory + OPE): $V(r) = -\alpha/r + c\cdot r^\gamma$,
and $\gamma = 2$ or $3$ depending on the parameters of the system
(see \cite{AkZa98,GuPoZaMW}).

Consider now the dual AHM with the action:
\beqn\label{AHM_action}
S=\int d^4x \big[
\frac{1}{4g^2} F^2_{\mu\nu} + 
\frac{1}{2} |(\partial - i B) \Phi|^2  + \\ \nonumber
+ \frac{1}{4} \lambda (|\Phi|^2-\eta^2)^2 \big],
\eeqn
where $B$ is the dual gauge field, $\Phi$ is the Higgs field carrying the
magnetic charge $g$. The Higgs field and the gauge field are massive:
$m_H^2= 2 \lambda \eta^2, m^2_V=g^2\eta^2$. The classical potential between
$+e$ and $-e$ charges (analogue of the quark -- antiquark pair) at large
distances is:

\beq
V(r) = \sigma_\infty\cdot r \qquad\qquad (r\to \infty)
\label{Vinf}
\eeq
The confining behavior is due to the dual Abrikosov - Nielsen -Olesen (ANO)
string which is formed between $+e$ and $-e$ charges. At small distances the
potential is Coulomb like: $V(r) = - \alpha/r, \,\, r \to 0$. We are
interested in the classical corrections to the Coulomb potential at
$r \ll m_H^{-1}, m_V^{-1}$ \cite{GuPoZa98}.
One has to impose the boundary condition, that is the vanishing of
the scalar field along a mathematically thin line connecting the magnetic
charges. This topological condition was
formulated in Ref.~\cite{tHo81} and implied in numerical
studies of the problem (see, e.g., \cite{alcock}).

This infinitely thin
line is nothing else but the Dirac string connecting the external
charges. By definition the energy of the Dirac string is zero. It is easy
to realize that the Dirac string cannot coexist with $\Phi\neq 0$.  Indeed,
if the Dirac string would be embedded into a vacuum with $\langle
\Phi\rangle\neq 0$ then its energy would again jump to infinity since there
is the term $1/2|\Phi|^2A_{\mu}^2$ in the action and $A_{\mu}^2\rightarrow
\infty$ for a Dirac string.
Therefore, even in the limit $r\to 0$ there is a deep well in
the profile of the Higgs field $\Phi$. This might cost energy which is
linear with $r$ even at small $r$.

Now we are coming to the next question, whether this mathematically thin
line realizes as a short {\it physical} string. By the physical string we
understand a stringy piece in the potential, $\sigma_0\cdot r$ at small $r$.
In other words, we are going to see whether the stringy boundary condition
implies a stringy potential.  To get the answer one solves the classical
equations of motion, the result is \cite{GuPoZa98}:

\beq\label{linearcorrection}
\lim_{r\to 0}V(r)~=~-{\alpha_M\over r}+\sigma_0\cdot r
+ ...
\eeq
with a non-vanishing slope $\sigma_0$.

The slope $\sigma_0$ depends smoothly on the value of $m_V/m_H$.
For example for $m_V=m_H$
the slope of the potential at $r\to 0$ is the same
as at $r\to \infty$. That is, within error bars:
\beq
\sigma_0~\approx~\sigma_{\infty} \label{ssimple}
\eeq
where $\sigma_{\infty}$ determines the value of the potential at large $r$,
\eq{Vinf}.

\section{Gauge boson propagator in the confining theory}

The simplest ``phenomenological'' theory of the quark confinement is
formulated in terms of the gluon propagator. Namely, one argues
\cite{West,BaBaZa88} that if the gluon propagator has a double pole in the
infrared region then the quark--antiquark potential gets a linear
attractive contribution at large distances and this implies the quark
confinement.  On the other hand, Gribov and Zwanziger ~\cite{Zw91} argued
that elimination of the gauge copies leads to a counter-intuitive vanishing
of the gluon propagator in the infrared region. Some analytical
studies~\cite{alagribov} confirm this behavior, while the 
others~\cite{nogribov} predict a singular gluon propagator in the infrared
limit. Recent numerical simulations in lattice
gluodynamics~\cite{ZwNumerical} show that the gluon propagator seems to
be finite at small momenta.

Below we consider the gauge boson propagator in dual AHM, that is the
abelian analogue of the gluon propagator. To simplify calculations we
consider the London limit, that is $\lambda \to \infty$ in \eq{AHM_action},
and $m_H \to \infty$. The final result remains unchanged if we relax the
condition on the coupling $\lambda$. To evaluate the propagator we utilize
the Zwanziger local field theory of electrically and magnetically charged
particles~\cite{Zw71}.  The theory contains two vectors potentials, namely
the gauge field $A_\mu(x)$ and the dual gauge field $B_\mu(x)$, which
interact covariantly with electric and magnetic currents, respectively. The
corresponding Lagrangian is:
\beqn \label{ZwInteraction}
\cL = \cL_{Zw}(A,B) 
+\, i \, e \, j^e_\mu(x) \, A_\mu (x) + \rule{10mm}{0mm}\\
+ i \, g \, j^g_\mu (x) \, B_\mu (x)\,, \nonumber
\eeqn
where $e$ ($g$) stands for the electric (magnetic) charge, and
$j^e_\mu$ ($j^g_\mu$) is the electric (magnetic) external current. The
Zwanziger Lagrangian $\cL_{Zw}$ is~\cite{Zw71}:
\beqn \label{ZwLagrangian}
\cL_{Zw} =
\frac{1}{2}(n\cdot[\diff\wedge A])^2 + \frac{1}{2}(n\cdot[\diff\wedge B])^2 +
\nonumber \\
+ \frac{i}{2}(n\cdot[\diff\wedge A])(n\cdot\dual{[\diff\wedge B]}) - \\ 
- \frac{i}{2}(n\cdot[\diff\wedge B])(n\cdot\dual{[\diff\wedge A]}) \,,\;\;\nonumber
\eeqn
where
$[A\wedge B]_{\mu\nu} =
A_\mu B_\nu - A_\nu B_\mu\,\, ,
\quad (n \cdot [A\wedge B])_\mu = n_\nu
(A\wedge B)_{\nu\mu}\,\, ,
\quad \dual{(G)}_{\mu\nu} = \frac{1}{2}
\varepsilon_{\mu\nu\lambda\rho} G_{\lambda\rho}$.
Inspite the Lagrangian~\eq{ZwInteraction} contains two gauge
fields, $A_\mu$ and $B_\mu$, there is only one physical particle (a massless
photon) in the spectrum~\cite{Zw71,BaRuSc75}. The Lagrangian
\eq{ZwInteraction} depends on an arbitrary constant unit vector $n_\mu$,
$n^2_\mu = 1$, while physical observables are insensitive to the direction
of $n$ provided the Dirac quantization condition, $e \cdot g = 2 \pi m$, $m
\in \Z$, is satisfied~\cite{Zw71}.

The theory of condensed monopoles interacting with an
external electric source $j^e$ (quark current) is described by the
following partition function~\cite{MaSu89}:
\beqn
\cZ[j^e]  = \int \dD A \dD B \dD \Phi
\, \exp\Bigl\{ - \int \dd^4 x \Bigl(
\cL_{Zw}(A,B) \nonumber\\
+ \frac{1}{2} {|(\partial + i g B) \Phi|}^2
+ \lambda {({|\Phi|}^2 - \eta^2)}^2 - 
i \, e \, j^e A \Bigr) \Bigr\}\, .\nonumber
\eeqn
We work in the Euclidean space and consider the London limit,
$\lambda \to \infty$.
The gauge field propagator is defined as:
\beqn \label{PropagatorDefinition}
D_{\mu\nu} (x,y) =
- \frac{1}{e^2} \, \frac{\delta^2}{\delta
j^e_\mu (x) \, \delta
j^e_\nu (y)} \frac{\cZ[j^e]}{\cZ[0]} {\Biggl|}_{j^e = 0}
\eeqn
To specify the propagator completely we choose the axial gauge $n_\mu A_\mu= 0$. 
The standard parameterization of the propagator in this gauge is \cite{BaBaZa88}:
\beqn
D^{\mathrm{axial}}_{\mu\nu} = \Bigl(\delta_{\mu\nu} + 
\frac{p_\mu n_\nu + p_\nu n_\mu}{{(p \cdot n)}} - \frac{p_\mu
p_\nu}{{(p \cdot n)}^2}\Bigr) F(p^2) -
\!\!\!\!\!\!\!\!\!\!\!\!\!\!\!\!\!\! \nonumber\\
- \frac{1}{{(p \cdot n)}^2}
\cdot (\delta_{\mu\nu} - n_\mu n_\nu)  G(p^2).
\label{baker}
\eeqn
Using methods of Refs.~\cite{BKTtoHiggs} the functions $F(p^2)$ and $G(p^2)$
can be expressed through the ANO string--string correlator \cite{ChPoZa99}:
\beqn
F(p^2)  =  \frac{1}{p^2 + m^2_V} \, 
\Bigl( 1 + \frac{m^4_V D^\Sigma (p^2)}{p^2 + m^2_V}\Bigr) \,
\label{f}\\
G(p^2)  =  - \frac{m^2_V}{p^2 + m^2_V}
\Bigl( 1 - m^2_V \frac{p^2 \, D^\Sigma (p^2)}{p^2 + m^2_V}\Bigr) \, ,
\label{g}
\eeqn
where
\beqn
D^{\Sigma}_{\mu \nu}(p)  =  - \frac{\eta^4 e^2}{(p^2 + m^2_V)(p \cdot n)} \cdot
\label{PropagatorString}  \\
\quad\qquad \cdot \int \frac{\dd^4 k}{{(2\pi)}^2} \frac{n_\alpha n_\beta {<\Sigma_{\mu\alpha}
(p) \Sigma_{\nu\beta} (-k)>}_\Sigma}{(k^2 +
m^2_V)(k \cdot n)} \, ,
\!\!\!\!\!\!\!\!\!\!\!\!\!\!\!\!\!\!\!\!\!\!\!\!\!\!\!\!\!\!\!\!\!
\nonumber
\eeqn
and the correlation
function of ANO strings \\ 
$<\Sigma_{\mu\alpha}(p) \Sigma_{\nu\beta} (-k)>$
is formally defined in~\cite{BKTtoHiggs}. 

The propagator \eq{baker} has singularities not only in
the $p^2$ plane but in the variable $(p\cdot n)$ as well.
The singular in $(p\cdot n)$ terms appear not only in the longitudinal
structures but also in front of $(\delta_{\mu\nu}-n_{\mu}n_{\nu})$.
It turns out that the singular structure
is due to the Dirac strings,
which are directed along vector $n_\mu$ entering the
Zwanziger Lagrangian \eq{ZwLagrangian}.
Note that Dirac strings become short strings
or ANO strings in the dual AHM.

Although expression (\ref{baker}-\ref{PropagatorString}) for the 
gluon propagator is exact, it
contains an unknown function $D^\Sigma (p^2)$. The infrared behavior of this 
function however can be qualitatively understood on physical grounds.  
Indeed, consider the string part \eq{PropagatorString} of the gluon
propagator \eq{baker}. This part corresponds to the following
process: the quark emits a gauge boson $A$ which transforms to a dual gauge
boson $B$ via coupling $<A \cdot B>$ which exists in the Zwanziger
Lagrangian \eq{ZwLagrangian}. The dual gauge boson transforms back to the
gauge boson after scattering on a closed ANO string world sheet $\Sigma$.
The intermediate string state is described by the function $D^\Sigma(p^2)$
and this state can be considered as a glueball state with the photon quantum
numbers $1^-$.

The behavior of the function $D^\Sigma(p^2)$ in the infrared region, $p\to 0$,
can be estimated as $D^\Sigma(p^2)=\frac{C}{p^2+M_{\mathrm{gl}}^2}+\dots$,
where $C$ is a dimensionless parameter and $M_{\mathrm{gl}}$ is the mass of 
the lowest $1^-$ glueball. The dots denote the contributions of heavier states.
Thus, according to eqs. (\ref{f},\ref{g}) in the infrared limit, $p\to 0$
\beqn
\label{infr}
F^{\mathrm{IR}}(p^2)  =  {1\over m^2_V} + {C\over M^2_{\mathrm{gl}}} + O(p^2) \\
G^{\mathrm{IR}}(p^2)  =  -1 +  O(p^2) \nonumber
\eeqn
Note that neither double, $1 \slash {(p^2)}^2$, nor ordinary,
$1\slash p^2$, pole
are exhibited by the gluon propagator (\ref{baker}-\ref{infr})
in the infrared limit. There are
gauge dependent $1/(p\cdot n)$ singularities which reflect
the presence of Dirac strings. Therefore our gauge boson propagator
does not vanish in the infrared either.
In view of this, the vanishing of the propagator at $p\to 0$
predicted in \cite{Zw91} seems to be specific for
the special gauge choice in QCD which leads to a nontrivial
Fundamental Modular Region.

\section{Relation of potential and propagator, violation of the operator 
product expansion.}

As it is mentioned above, the existence of short strings is manifested
also through breaking of the standard OPE. The physical reason of this
breaking  is that strings have extra dimensionality compared to particles, 
they are not point--like objects.
The another manifestation of this problem is the violation of the
standard relation between the potential and the propagator. Let us try to
reconstruct the classical Coulomb potential between (heavy) electric
charge--anticharge and between magnetic charge--anticharge in the dual AHM
for $r \ll m_V^{-1},\, m_H^{-1}$.
For magnetic charges the potential can be reconstructed in the
standard way from the propagator $\tilde{D}_{\mu\nu}(q^2)$ in the momentum
space,
\beq V(r)=\int d^3 q \; e^{i{\bf q\cdot r}}\;
\tilde{D}_{00}({\bf q}^2) \, .
\label{aa}\eeq
In order to have the standard notations we use here $q$ in the Minkovsky
region, $q^2 = - p^2$ ({\it cf.} eq. \eq{baker} -- \eq{PropagatorString}).
For $q$ in Euclidean region and much
larger than the mass parameters, the propagator $\tilde{D}_{\mu\nu}(q^2)$
can be evaluated by using the OPE.  Restriction to the classical
approximation implies that loop contributions are not included.  However,
vacuum fields which are soft on the scale of ${\bf q}^2$ can be consistently
accounted for in this way (for a review of see, e.g., \cite{novikov}).
Modulus longitudinal terms, we have:

\beqn
\label{prop} 
\tilde{D}_{\mu\nu}(q^2) = \delta_{\mu\nu} 
\left(
{1\over q^2}+{1\over q^2} g^2 \langle \Phi^2 \rangle {1\over q^2}+ \right. \nonumber \\
\left. {1\over q^2}g^2\langle \Phi^2\rangle { 1\over q^2}g^2\langle \Phi^2
\rangle  {1\over q^2}+... \right)
= {\delta_{\mu\nu}\over q^2-m_V^2}.
\eeqn
Thus, one uses first the general OPE assuming $|q^2|\gg g^2\Phi^2$ then
substitutes the vacuum expectation of the Higgs field $\Phi$ and upon
summation of the whole series of the power corrections reproduces the
propagator of a massive particle ($<\Phi^2> = \eta^2$, $m^2_V = g^2
\eta^2$). This propagator according to \eq{aa} corresponds to Yukawa
potential, or to Coulomb potential at small distances.

The photon exchange between electric charges (analogues of
quarks) corresponds to propagator \eq{baker}. At the classical level we
should neglect the contribution of virtual strings, $D^\Sigma(p^2) = 0$,
and:
\beqn
D_{\mu\nu}(q^2) = {1\over q^2-m_V^2}(\delta_{\mu\nu}-
{1\over (qn)}(q_{\mu}n_{\nu}+q_{\nu}n_{\mu}) 
\!\!\!\!\!\!\!\!\!\!\!\!\!\!
\nonumber \\
+ {q_{\mu}q_{\nu}\over(qn)^2}+
{m_V^2\over (qn)^2}(\delta_{\mu\nu}n^2-n_{\mu}n_{\nu})) .
\label{wrong}
\eeqn

If the potential energy is given by the Fourier transform of (\ref{wrong})
then its dependence on $n_{\mu}$ is explicit.
Thus it is impossible to get the Coulomb potential by the Fourier transform
(\ref{aa}), we addressed this problem in Ref. \cite{GuPoZa97}.

Note that Eq. (\ref{wrong}) immediately implies that the standard OPE
does not work any longer on the level of $q^{-2}$ corrections.
Indeed, choosing $q^2$ large and negative does not guarantee now that
the $m_V^2$ correction is small since the factor $(qn)^2$ in the
denominator may become zero.

The reason for the breaking of the
standard OPE is that even at short distances the dynamics of the short
strings should be accounted for explicitly. In particular, in the
classical approximation the string lies along the straight line
connecting the magnetic charges and affects the solution through the
corresponding boundary condition, see above.
The OPE is valid as far as the probe particles do
not change the vacuum fields drastically and the unperturbed vacuum
fields are a reasonable zero-order approximation.  In our case,
however, the Higgs field is brought down to zero along the string and
this is a nonperturbative effect.

\section{Short strings in QCD}

Now we briefly review the possible manifestations of short strings in 
numerical simulations of lattice gluodynamics.

{\it (i)} The lattice simulation \cite{bali2} do
not show any change in the slope of the full $Q\bar{Q}$ potential as the
distances are changed from the largest to the smallest ones 
where Coulombic part becomes dominant.
Thus eq.~\eq{ssimple} is presumably satisfied. Moreover, it is known from 
phenomenological
analysis and from the calculations on the lattice \cite{Suz88,Blimit,BSS98} 
that the realistic QCD corresponds to the case $\kappa\approx 1$ where
$\kappa =m_H/m_V$. It is remarkable that, as is mentioned above, the AHM in
the classical approximation also results in the relation (\ref{ssimple}) for
$\kappa\approx 1$.

{\it (ii)} Addressing the question of linear corrections at small distances
one tries to explicitly substruct the perturbative contribution from
the full $Q\bar{Q}$ potential. It appears that the remaining part is well
parameterized by linear function and the 
extrapolation of the lattice data
to very small distances gives large value for $\sigma_0$,
$\sigma_0 \approx 6 \sigma_\infty$ \cite{bali3}.

{\it (iii)} There exist lattice measurements of fine splitting of the
$Q\bar{Q}$ levels as a function of the heavy quark mass. The
Voloshin-Leutwyler\cite{VL} picture results in a particular pattern of the
mass dependence of this splitting. Moreover, these predictions are very
different from the predictions based on adding a linear part to the
Coulomb potential (Buchmuller-Tye potential \cite{BuTy81}).
The results of most advanced measurements of this type \cite{Fi98} are in 
favor of the linear (\ref{linearcorrection}-\ref{ssimple})correction to 
the potential at short distances.

({\it iv}) There is an evidence that the
nonperturbative fluctuations on the lattice responsible for the confinement
can be identified with the so called P vortices \cite{Fa99}.
If one measures the quark potential due only to these vortices, the
numerical results seem  to indicate that the slope of the potential is the
same at large and at small distances, see \cite{Deb97}.

({\it v}) The  dimensional analysis show that the linear correction at 
short distances to the potential $V(r)$ produced by short string corresponds
to $1/Q^2$ correction in OPE. The standard leading correction 
is only $1/Q^4$, which corresponds to the operator $<Tr F_{\mu\nu}^2>$. 
Numerical data for the
expectation value of the plaquette minus perturbation theory contribution
show $1/Q^2$ behavior \cite{March}.

At present the systematical errors in numerical calculations are large,
especially the subtraction of perturbation theory contribution in items {\it 
(ii)} and {\it (v)} is not a well established procedure, nevertheless it 
seems that there are several independent facts showing the existence of the 
short strings in lattice gluodynamics.

Now we discuss two effects which can be manifestation of the short strings
in QCD.

({\it i}) Analytical studies of the Bethe-Salpeter equation and comparison
of the results with the charmonium spectrum data favor a
nonvanishing linear correction to the potential at short distances
\cite{BaMo99}.

({\it ii}) The most interesting manifestation of short strings is $1/Q^2$
corrections to the standard OPE for current--current correlation function
$\Pi(Q^2)$ \cite{AkZa98,CNZ99,GuPoZaMW}. It
is impossible to calculate the coefficient of $1/Q^2$ corrections, and in
ref. \cite{CNZ99} it was suggested to simulate this correction by tahyonic
gluon mass. The Yukawa potential with an imaginary mass has the linear
attractive piece at small distances as it is induced by short strings. The
use of the gluon propagator with the imaginary gluon mass 
($m^2_g = -0.5 \mbox{ Gev}^2$) 
unexpectively well explains the different behavior of $\Pi(Q^2)$ in
different channels. It is very important to perform the accurate
calculations of various correlators $\Pi(Q^2)$ on the lattice to check the
model. There are theoretical schemes \cite{ImT} which imitate the tachyonic
gluon mass.

\section*{Acknowledgements}
We are acknowledging thankfully discussions with A.~Leonidov, V.A.~Novikov,
L.~Stodolsky, A.I.~Vainshtein, M.I.~Vysotsky. Work of M.N.C., F.V.G. and
M.I.P.  was partially supported by grants INTAS-RFBR-95-0681, RFBR 99-01230a
and INTAS 96-370.

\end{document}